\documentclass[12pt,a4paper]{article}
\usepackage{amsmath,amssymb,slashed,cancel}
\usepackage{cite,tabularx}
\usepackage{subcaption}
\usepackage{graphicx,color}
\usepackage[normalem]{ulem}
\usepackage{mathtools}
\usepackage{enumerate}
\usepackage{multirow}
\usepackage{bm}
\usepackage{braket}
\allowdisplaybreaks

\usepackage[height=8.85in,width=6.4in]{geometry}

\newcommand\package[2][\relax]{\texttt{#2\ifx#1\relax\relax\relax\else\,\linebreak[0]#1\fi}}

\numberwithin{equation}{section} % Eq.(Sec.eq.)
 
\def\beq#1\eeq{\begin{align}#1\end{align}}

\definecolor{BlueViolet}{rgb}{0.2, 0.00, 0.7}
\definecolor{Blue}{rgb}{0.15, 0.00, 0.9}
\usepackage[colorlinks=true,linkcolor=Blue,citecolor=Blue,urlcolor=BlueViolet]{hyperref}

\begin{document}
\begin{titlepage}
\setcounter{page}{0} % to set page=0 of the title-page

\begin{center}

\hfill {\tt STUPP-22-258}\\

\vskip .55in

\begingroup
\centering
\large\bf Asymmetric Mediator in Scotogenic Model
\endgroup

\vskip .4in

\renewcommand{\thefootnote}{\fnsymbol{footnote}}
{
Kento Asai$^{(a,b,c)}$\footnote{
  \href{mailto:kento@icrr.u-tokyo.ac.jp}
  {\tt kento@icrr.u-tokyo.ac.jp}},
Yuhei Sakai$^{(c)}$\footnote{
  \href{mailto:sakai@krishna.th.phy.saitama-u.ac.jp}
  {\tt sakai@krishna.th.phy.saitama-u.ac.jp}},
Joe Sato$^{(b,c)}$\footnote{
  \href{mailto:sato-joe-mc@ynu.ac.jp}
  {\tt sato-joe-mc@ynu.ac.jp}},
Yasutaka Takanishi$^{(b,c)}$\footnote{
  \href{mailto:yasutaka@krishna.th.phy.saitama-u.ac.jp}
  {\tt yasutaka@krishna.th.phy.saitama-u.ac.jp}},\\ and
Masato Yamanaka$^{(b)}$\footnote{
  \href{mailto:yamanaka-masato-xc@ynu.ac.jp}
  {\tt yamanaka-masato-xc@ynu.ac.jp}}
}

\vskip 0.4in

\begingroup\small
\begin{minipage}[t]{0.9\textwidth}
\centering\renewcommand{\arraystretch}{0.9}
{\it
\begin{tabular}{c@{\,}l}
$^{(a)}$
& Institute for Cosmic Ray Research (ICRR), The University of Tokyo, Kashiwa,\\
& Chiba 277--8582, Japan \\[2mm]
$^{(b)}$
& Department of Physics, Faculty of Engineering Science, Yokohama National University, \\
& Yokohama 240--8501, Japan \\[2mm]
$^{(c)}$
& Department of Physics, Faculty of Science, Saitama University, Saitama 338--8570, \\ 
& Japan \\
\end{tabular}
}
\end{minipage}
\endgroup

\end{center}

\vskip .4in

\begin{abstract}\noindent

The scotogenic model is the Standard Model (SM) with $Z_2$ symmetry
and the addition of $Z_2$ odd right-handed Majorana neutrinos and SU(2)$_L$
doublet scalar fields. 
We have extended the original scotogenic model by an
additional $Z_2$ odd singlet scalar field that plays a role in dark matter. 
In our model, the asymmetries of the lepton and $Z_2$ odd doublet scalar are simultaneously produced through CP-violating right-handed neutrino decays. 
While the former is converted into baryon asymmetry through the sphaleron process, the latter is relaid to the DM density through the decay of SU(2)$_L$ doublet scalar that is named ``asymmetric mediator''. 
In this way, we provide an extended scotogenic model that predicts  
the energy densities of baryon and dark matter being in the same order of magnitude,  
and also explains the low-energy neutrino masses and mixing angles.

\end{abstract}
\end{titlepage}
%%%
\setcounter{page}{1}
\renewcommand{\thefootnote}{\#\arabic{footnote}}
\setcounter{footnote}{0}

%%%%%%%%%%%%%%%%%%%%%%%%%%%%%%%%%%%%%%%%%%%%%%%%%%%%%%%%
%\begingroup
%\renewcommand{\baselinestretch}{1} % this command affects only in this "\begingroup-\endgroup"
%\setlength{\parskip}{2pt}          % this command affects only in this "\begingroup-\endgroup"
%\hrule
%\tableofcontents
%\vskip .2in
%\hrule
%\vskip .4in
%\endgroup

%%%%%%%%%%%%%%%%%%%%%%%%%%%%%%%%%%%%%%%%%%
%     Introduction
%%%%%%%%%%%%%%%%%%%%%%%%%%%%%%%%%%%%%%%%%%
\section{Introduction}
\label{sec:introduction}
The existence of dark matter (DM) and the non-zero value of the baryon
asymmetry of the universe (BAU) are long-standing unsolved puzzles in
the standard theory of cosmology and particle physics.  In fact there
is no candidate for the DM particle in the standard model (SM), and
accordingly, various particles have been suggested and studied
intensively for the DM in particle theories beyond the SM (BSM). One
of the most promising candidates of DM is the so-called weakly
interacting massive particle (WIMP), and the relic abundance of the DM
can be calculated by its annihilation cross-section for the thermal
freeze-out scenario.

The other problem, namely, BAU is positively realized through the
mechanism of leptogenesis~\cite{Fukugita:1986hr}.
Lepton asymmetry is generated by the CP-violating decay
of the right-handed neutrinos. Then this lepton
asymmetry is converted into baryon asymmetry through the sphaleron
process. It is obvious that the amount of the produced baryon
asymmetry is determined by the masses of the right-handed neutrinos
and Yukawa couplings in this thermal leptogenesis scenario.
%%%%%%%%%%%%%%%%%%%%%%%%%%%%%%%%%%%%%%%%%%

The relic abundance of the DM and baryon are measured by the 
Planck observations of the cosmic microwave background (CMB), and the current 
values are given in the following~\cite{Planck:2018vyg}:
\begin{align}
\label{eq:OmegaDM-OmegaB}
    \Omega_{\rm DM} h^2 &= 0.120 \pm 0.001~, \\
    \Omega_{\rm B} h^2 &= 0.0224 \pm 0.0001 ~,
\end{align}
where we express the Hubble constant $h$ in units of 100 km/s/Mpc.
Surprisingly, these two abundances are strikingly similar as $\Omega_{\rm DM} / \Omega_{\rm B} \approx 5$, although the DM and BAU are independently produced through different processes in general. 
This coincidence of these relic abundances implies the existence of mechanisms that link the productions of the DM and BAU together.
%%%%%%%%%%%%%%%%%%%%%%%%%%%%%%%%%%%%%%%%%%

Asymmetric dark matter (ADM)~\cite{Nussinov:1985xr,Barr:1990ca,Barr:1991qn,Dodelson:1991iv,Kaplan:1991ah,Kuzmin:1996he,Foot:2003jt,Foot:2004pq,Hooper:2004dc,Kitano:2004sv,Gudnason:2006ug,Kaplan:2009ag,Davoudiasl:2012uw,Petraki:2013wwa,Zurek:2013wia} is one of the frameworks where the coincidence between the relic abundances of the DM and baryon is realized.
In this framework, an asymmetry of the DM and anti-DM number densities is produced in the early universe.
As the universe cools, the symmetric component annihilates into the SM particles, and then the remaining asymmetry component explains the observed relic abundance of the DM.
Generations of the DM asymmetry are roughly classified into two types.
One is the sharing mechanism that the asymmetry related with BAU in the SM sector is firstly generated, and then the produced asymmetry is shared between the DM and SM sectors through some interactions.
The other is the cogenesis mechanism in which the asymmetries of the matter and DM are generated simultaneously. 

In this article, we focus on the cogenesis mechanism in the scotogenic model.
The scotogenic model~\cite{Ma:2006km} is one of the seesaw models~\cite{Minkowski:1977sc,Yanagida:1979as,Gell-Mann:1979vob,Mohapatra:1979ia}.
In scotogenic model, right-handed neutrinos and a neutrino-philic inert SU(2)$_L$ doublet scalar are introduced, and these fields transform odd under an exact $Z_2$ symmetry.
The masses of the light neutrinos are generated through an one-loop diagram.
Moreover, the lightest $Z_2$ odd field is stable and can be a candidate for the DM.
However, this $Z_2$ odd scalar can not be a DM in the context of the ADM.
This is because the mass of the DM should be the same order of magnitude as that of proton, and such a light DM which interacts with the weak gauge bosons is strongly constrained by the requirement that neutron stars do not gravitationally collapse into black holes~\cite{Goldman:1989nd,Gould:1989gw,Kouvaris:2007ay,Kouvaris:2010vv,deLavallaz:2010wp,McDermott:2011jp,Kouvaris:2011fi,Guver:2012ba,Bell:2013xk,Bramante:2013hn,Bramante:2013nma,Kouvaris:2013kra,Bramante:2014zca,Ilie:2020vec,Kouvaris:2018wnh,Gresham:2018rqo,Garani:2018kkd}.

From these reasons we introduce an additional $Z_2$ odd singlet real scalar as the DM in the scotogenic model.
In this new model, the SU(2)$_L$ doublet scalar plays a role of the mediator which links the DM and right-handed neutrinos and relays the asymmetry to the DM. 
Firstly, the CP-violating decays of the right-handed neutrinos generate the same amount of the lepton and mediator asymmetries simultaneously.
After annihilation of the symmetric component of the mediators, the asymmetric component decays into the DM, and then the mediator asymmetry converts into the relic abundance of the DM.
Thus, the same order of number densities of the baryon and DM are realized.~\footnote{There is another model which realizes the coincidence between DM abundance and baryon asymmetry in frameworks of the scotogenic model~\cite{Borah:2018uci,Borah:2019epq}.
In these papers, the lepton asymmetry is generated through annihilation and coannihilation of dark sector particles.}

%%%%%%%%%%%%%%%%%%%%%%%%%%%%%%%%%%%%%%%%%%%
This article organized as follows. Next section 
we review the scotogenic model and neutrino parameters.
In Sec.~\ref{sec:cogenesis}, we discuss the leptogenesis and DM production in the scotogenic model with a real singlet scalar DM.
In Sec.~\ref{sec:result}, we show the parameter region where the model in this paper explains the observed baryon asymmetry, DM density, and neutrino mixing parameters simultaneously.
Finally, our conclusions are discussed in Sec.~\ref{sec:summary}.

%%%%%%%%%%%%%%%%%%%%%%%%%%%%%%%%%%%%%%%%%%
%     Models
%%%%%%%%%%%%%%%%%%%%%%%%%%%%%%%%%%%%%%%%%%
\section{Scotogenic Model with Singlet Scalar Dark Matter}
\label{sec:model}
In this section, we introduce a new scotogenic model
by adding a real singlet scalar field. The matter contents of the
original scotogenic model are of the SM plus three right-handed neutrinos $N_i~(i = 1,2,3)$ and an inert doublet scalar $\eta$. The SM
fields are even under a discrete $Z_2$ symmetry but non-SM fields: the
right-handed neutrinos $N_i (i = 1,2,3)$, an inert doublet scalar
$\eta$, and a single scalar $\sigma$, are odd under this symmetry. In
Tab.~\ref{tab:fields}, the matter contents of our model are summarized.

It is important to note that a singlet scalar $\sigma$ plays a role of
DM in our model~\footnote{The scalar field $\sigma$ may be
also a complex scalar field. There is no difference between
these choices except for the degree of freedom.}. As will be mentioned 
below, this field is the lightest particle 
among the $Z_2$-odd fields, and therefore the stability of the dark 
matter is guaranteed.

%%% Table %%%%%%%%%%%%%%%%%%%%%%%%%%%%%%%%%%%
\begin{table}[t]
\begin{center}
\begin{tabular}{|ccccccc|} \hline
\multirow{2}{*}{field} & \multicolumn{3}{c}{fermion} & \multicolumn{3}{c|}{scalar} \\
& $L$ & $e_R^{}$ & $N$ & $H$ & $\eta$ & $\sigma$ \\ \hline
SU(2)$_L$ & $\mathbf{2}$ & $\mathbf{1}$ & $\mathbf{1}$ & $\mathbf{2}$ & $\mathbf{2}$ & $\mathbf{1}$ \\
$Z_2$ & $+$ & $+$ & $-$ & $+$ & $-$ & $-$ \\ \hline
\end{tabular}
\caption{Matter contents of the extended version of scotogenic model.}
\label{tab:fields}
\end{center}
\end{table}
%%%%%%%%%%%%%%%%%%%%%%%%%%%%%%%%%%%%%%%%%%%%
%

The SM left-handed lepton doublet, the right-handed charged lepton,
and the Higgs doublet scalar are denoted by $L$, $e_R^{}$, and $H$,
respectively. Under this setup, the Lagrangian relative to the
neutrinos and scalar potential is given by
\begin{align}
\label{eq:lagrangial}
   \mathcal{L} 
   \supset &\,- h_{\alpha i} \bar{L}_\alpha \tilde{\eta} N_i + \frac{1}{2} M_i \bar{N}_i N_i^c + {\rm H.c.}~, \\
   V(H,\eta,\sigma) =
   &\,\mu_H^2 |H|^2 + m_\eta^2 |\eta|^2 + \frac{1}{2} m_\sigma^2 \sigma^2 + \frac{1}{2} \lambda_1 |H|^4 + \frac{1}{2} \lambda_2 |\eta|^4 + \frac{1}{2} \lambda_3 \sigma^4 + \lambda_4 |H|^2 |\eta|^2 \nonumber \\
   &\,+ \lambda_5 |H^\dag \eta|^2 + \lambda_6 |H|^2 \sigma^2 + \lambda_7 |\eta|^2 \sigma^2 + \frac{1}{2} \left[ \lambda_8 (H^\dag \eta)^2 + {\rm H.c.} \right] \nonumber \\
   &\,+ \frac{1}{\sqrt{2}} \left[ \mu \sigma (H^\dag \eta) + {\rm H.c.} \right]~,
\end{align}
where $\alpha~ (i)$ denotes the index of the flavor (mass) eigenstates, $M_i$ represents the mass eigenvalue of the heavy neutrino $N_i$, $\tilde{\eta} \equiv i \sigma_2 \eta^*$, and $\mu_H^2$ is negative.
All the parameters in the scalar potential can be chosen real without loss of generality.
Only the SM Higgs acquires a nonzero vacuum expectation value (VEV), and the other scalars do not.

As we will show in the following section, our scenario works under the condition 
that $m_\eta$ ($\mu$) is much higher (lower) than the electroweak scale. Therefore 
the mixing between CP-even neutral components of $\eta$ and $\sigma$ is negligible although $\mu$ plays an important role in the dark matter production and should not be zero.
Additionally, we assume $\lambda_6, \lambda_7 \ll 1$ to avoid constraints from direct detection experiments and thermalization of the DM in the early universe. After the SM Higgs acquires a nonzero VEV, the masses of the charged, CP-even, and CP-odd components of the inert doublet scalar, $\eta = (\eta^+, \eta^0)^T$ with $\eta^0 = (\eta_R^{} + i \eta_I^{})/\sqrt{2}$, split and are given by
\begin{align}
    m_{\eta^+}^2 =
    &\,m_\eta^2 + \frac{1}{2} \lambda_4 v^2~, \\
    m_{\eta_R^{}}^2 \simeq
    &\,m_\eta^2 + \frac{1}{2} (\lambda_4 + \lambda_5 + \lambda_8) v^2~, \\
    m_{\eta_I^{}}^2 \simeq
    &\,m_\eta^2 + \frac{1}{2} (\lambda_4 + \lambda_5 - \lambda_8) v^2~,
\end{align}
where $v$ is the VEV of the SM Higgs field.
That of the singlet scalar DM is given by
\begin{align}
    m_{\rm DM}^2 \simeq  m_\sigma^2 + \frac{1}{2} \lambda_6 v^2~.
\end{align}

\begin{figure}[t]
\centering
\includegraphics[width=0.6\textwidth]{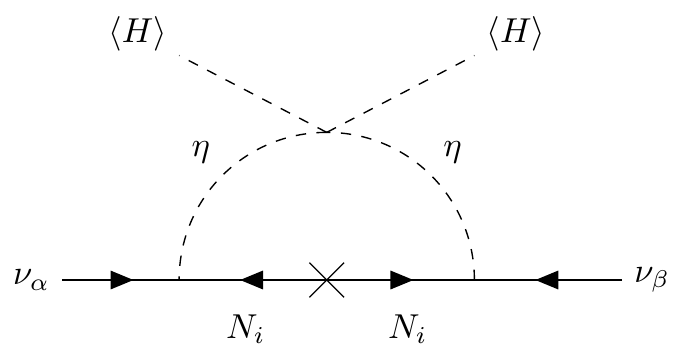}
\caption{One-loop diagram that generates the neutrino masses.}
\label{fig:neutrino-mass}
\end{figure}

The neutrino masses are radiatively generated as shown in Fig.~\ref{fig:neutrino-mass}.
\footnote{There is another contribution to the active neutrino mass by $\sigma$. 
Note that effectively $\lambda_8$ term is induced by the
exchange of $\sigma$ with the coupling $\mu \sigma (H^\dag \eta)$.
It is negligible because $\lambda_8 \gg {({\mu}/{m_\sigma})}^2$ in this model, and we do not
discuss it here.
}
%
%%% Figure %%%%%%%%%%%%%%%%%%%%%%%%%%%%%%%%%%%
%%%%%%%%%%%%%%%%%%%%%%%%%%%%%%%%%%%%%%%%%%%%
%
The mass matrix of the active neutrinos are obtain as
\begin{align}
\label{eq:Mnu}
    (\mathcal{M}_\nu)_{\alpha \beta} =
    \sum_i \frac{h_{\alpha i}^* h_{\beta i}^*}{32 \pi^2} M_i \left[ \frac{m_{\eta_R^{}}^2}{m_{\eta_R^{}}^2 - M_i^2} \ln{\frac{m_{\eta_R^{}}^2}{M_i^2}} - \frac{m_{\eta_I^{}}^2}{m_{\eta_I^{}}^2 - M_i^2} \ln{\frac{m_{\eta_I^{}}^2}{M_i^2}} \right]~.
\end{align}
In addition to the assumption that $m_\eta$ is much higher than the electroweak scale, we assume further that the mass of the right-handed neutrinos are much heavier than $m_{\eta_R^{}}$ and $m_{\eta_I^{}}$. 
Thus, for $m_{\eta_{R,I}^{}}^{} \ll M_i$ and $\lambda_8 v^2 \ll m_\eta^2$, the mass matrix of the active neutrinos can be approximated as follows:
\begin{align}
\label{eq:Mnu-approx}
    (\mathcal{M}_\nu)_{\alpha \beta} \simeq
    \frac{\lambda_8 v^2}{32 \pi^2} \sum_i \frac{h_{\alpha i}^* h_{\beta i}^*}{M_i} \left[ \ln{\frac{M_i^2}{m_0^2} - 1} \right]~,
\end{align}
where $m_0^2 \equiv (m_{\eta_R^{}}^2 + m_{\eta_I^{}}^2) / 2$\,.

For convenience, we introduce the Casas-Ibarra (CI) parametrization~\cite{Casas:2001sr}, following Ref.~\cite{Hugle:2018qbw} in which the leptogenesis scenario in the scotogenic model is discussed.
The mass matrix for the light neutrinos is rewritten by the diagonal matrix $\mathcal{D}_\Lambda^{}$ in the following way:
\begin{align}
\label{eq:Mnu_Lambda}
   \mathcal{M}_\nu &= h^* \mathcal{D}_\Lambda^{-1} h^\dag~, \\
\label{eq:Lambda}
   (\mathcal{D}_\Lambda^{})_{ii} &= \frac{2\pi^2}{\lambda_8} \xi_i \frac{2 M_i}{v^2} \equiv \Lambda_i~,
\end{align}
with
\begin{align}
\label{eq:xi}
   \xi_i \equiv \left\{ \frac{1}{8} \frac{M_i^2}{m_{\eta_R^{}}^2 - m_{\eta_I^{}}^2} \left( \frac{m_{\eta_R^{}}^2}{m_{\eta_R^{}}^2 - M_i^2} \ln{\frac{m_{\eta_R^{}}^2}{M_i^2}} - \frac{m_{\eta_I^{}}^2}{m_{\eta_I^{}}^2 - M_i^2} \ln{\frac{m_{\eta_I^{}}^2}{M_i^2}} \right) \right\}^{-1}~.
\end{align}
For the hierarchical masses structure between the right-handed neutrinos and inert scalars ($m_\eta \ll M_i$) and the small scalar four-point coupling ($\lambda_8 v^2 \ll m_\eta^2$), the parameters $\xi_i$ can be approximated as 
\begin{align}
\xi_i \approx \frac{8}{\left[ \ln{(M_i^2/m_0^2)} - 1 \right]} ~.
\end{align}

The mass matrix for the light neutrinos can be diagonalized by a unitary matrix called Pontecorvo-\-Maki-\-Nakagawa-\-Sakata (PMNS) matrix $U$~\cite{Pontecorvo:1967fh,Pontecorvo:1957cp,Pontecorvo:1957qd,Maki:1962mu} as 
\begin{align}
U^T \mathcal{M}_\nu U = {\rm diag}(m_1, m_2, m_3) \equiv D_\nu
\end{align}
Note that we follow the convention of the Particle Data Group~\cite{Zyla:2020zbs}, and it is different from that of Ref.~\cite{Hugle:2018qbw}. The Yukawa couplings are written as follows:
\begin{align}
\label{eq:Yukawa}
   h_{\alpha i} = \left( U\, D_\nu^{\frac{1}{2}}\, R^\dag\, D_{\Lambda}^{\frac{1}{2}} \right)_{\alpha i}~,
\end{align}
where $R$ is an arbitrary complex orthogonal matrix satisfying $R R^T = 1$.

%%%%%%%%%%%%%%%%%%%%%%%%%%%%%%%%%%%%%%%%%%
%     Cogenesis
%%%%%%%%%%%%%%%%%%%%%%%%%%%%%%%%%%%%%%%%%%
\section{Cogenesis Mechanism in the Scotogenic Model}
\label{sec:cogenesis}

%%% Figure %%%%%%%%%%%%%%%%%%%%%%%%%%%%%%%%%%%
\begin{figure}[t]
\centering
\includegraphics[width=0.8\textwidth]{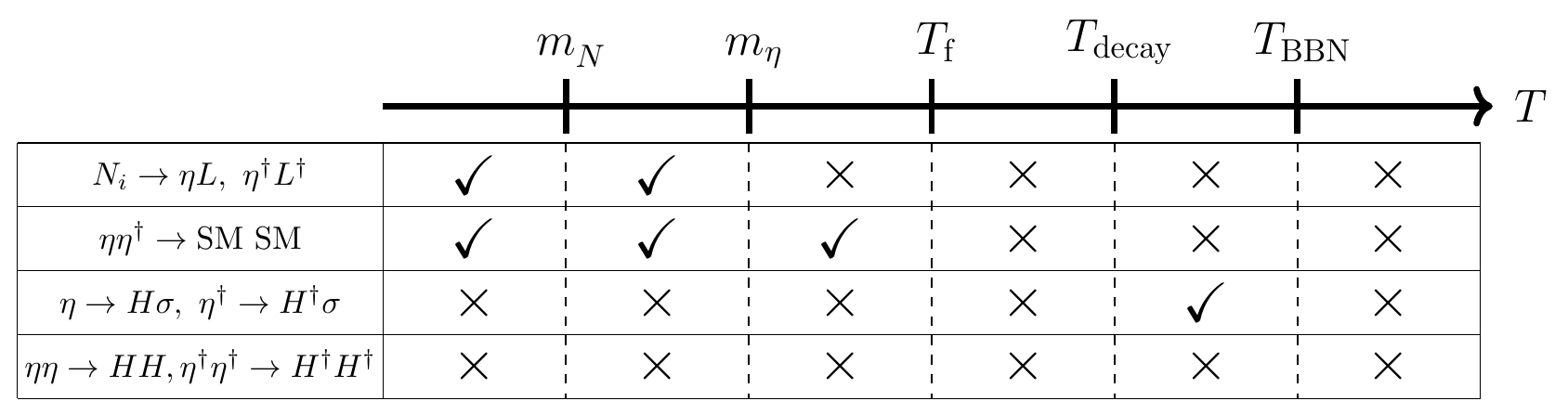}
\caption{
Summary of our scenario for generating the baryon asymmetry and DM relic abundance.
The mass of right-handed neutrino (mediator) is denoted by $m_N (m_\eta)$.
$T_{\rm f}, T_{\rm decay}$, and $T_{\rm BBN}$ is the freeze-out temperature of the mediator annihilation, that when the mediator decay gets active, and that when the Big-Bang Nucleosynthesis (BBN) begins, respectively.
Check ($\checkmark$) and cross ($\times$) marks represent that the 
corresponding process becomes important and ineffective, respectively.}
\label{fig:summary}
\end{figure}
%%%%%%%%%%%%%%%%%%%%%%%%%%%%%%%%%%%%%%%%%%%%

In this section, we discuss the generation of the baryon asymmetry and DM through the cogenesis mechanism in our extended scotogenic model.
Firstly, we summarize the story of the generation of the DM and baryon asymmetries in the extended scotogenic model. Our setup is displayed in Fig.~\ref{fig:summary}.

\begin{figure}[t]
\centering
\includegraphics[width=1.0\textwidth]{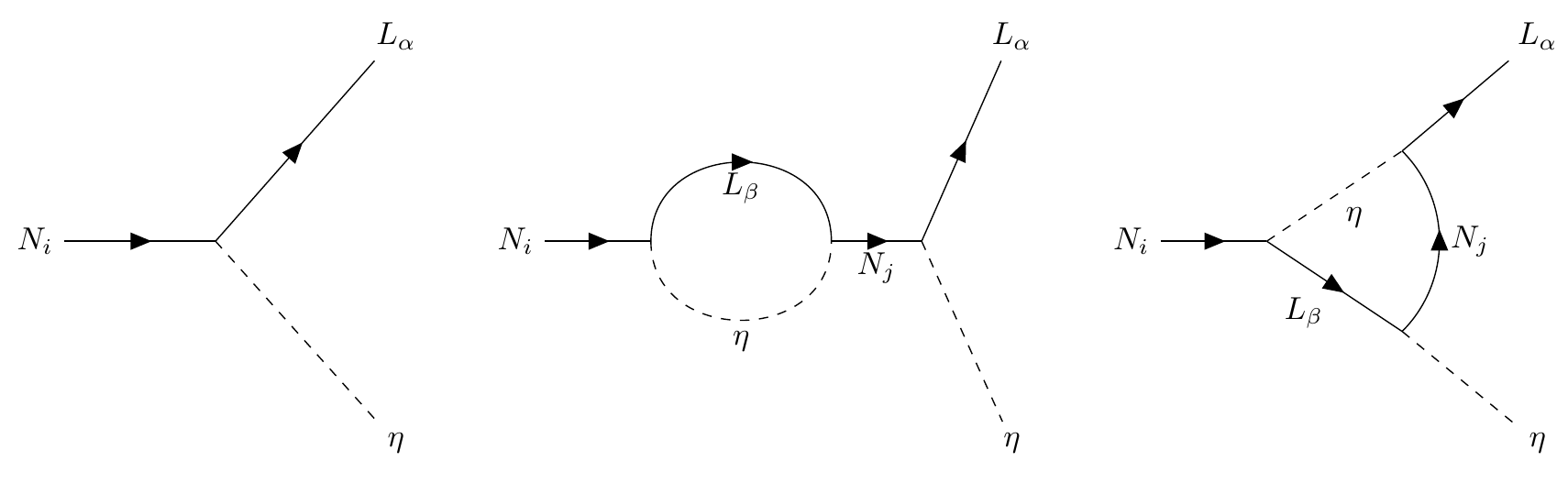}
\caption{Feynman diagrams contributing to the asymmetries of the baryon and inert doublet scalar.}
\label{fig:leptogenesis}
\end{figure}

In the early universe, the right-handed neutrinos are thermally produced in the SM thermal plasma.
After the temperature gets lower than the mass of the lightest right-handed neutrino $N_1$~\footnote{
In this article, we assume that the asymmetries of the baryon and inert doublet scalar are dominantly generated by the decay of the lightest right-handed neutrino $N_1.$}, the decay process of $N_1$ becomes out-of-equilibrium.
Then, the asymmetries of the lepton and mediator are generated by the CP-violating decays of $N_1$ as shown in Fig.~\ref{fig:leptogenesis}.
%
%%% Figure %%%%%%%%%%%%%%%%%%%%%%%%%%%%%%%%%%%
%%%%%%%%%%%%%%%%%%%%%%%%%%%%%%%%%%%%%%%%%%%%
%
We must notice that the interaction $\eta \eta \leftrightarrow H H$ should become out-of-equilibrium to avoid the wash-out of the mediator asymmetry. 
Therefore the coupling $\lambda_8$ should be small, but on the other hand, sufficiently too small $\lambda_8$ spoils the generation of the neutrino masses by the scotogenic seesaw mechanism.

%%%%%%%%%

After the temperature drops lower than the mediator mass, they quickly annihilate into the SM fields through the SU(2)$_L$ interaction.
The mediator asymmetry is generated in the same process as that of the lepton, and the annihilation process does not change the difference of the number density between $\eta$ and $\eta^\dag$: $n_{\Delta \eta}^{} \equiv n_\eta - n_{\eta^\dag}$ with $n_{\eta} \equiv n_{\eta^0} + n_{\eta^+}$ and $n_{\eta^\dag} \equiv n_{\eta^{0*}} + n_{\eta^-}$, as long as $\lambda_8$ is small enough to neglect the CP violating annihilation, $\eta \eta \to H H\, (\eta^\dag \eta^\dag \to H^\dag H^\dag)$. 
During the annihilation, $n_{\Delta \eta}^{}$ is, therefore, equal to the lepton asymmetry $n_{\Delta L}^{}$ if the decay rate of $\eta$ and $\eta^\dag$ is less than the Hubble parameter $H$.
After falling out of equilibrium of the annihilation process, the density of $\eta$ becomes much smaller than that of $\eta^\dag$, and the hierarchy of the number densities is realized as $|n_{\Delta \eta}^{}| \simeq n_{\eta^\dag} \gg n_{\eta}$.
Subsequently, the decays of the remaining $\eta^\dag$ start at the temperature $T_{\rm dec}$, and then, $n_{\Delta \eta}^{}$ is converted into the DM number density $n_{\rm DM}^{}$. 
Thus, the number density of the DM has the same order as those of the SM lepton and baryon.
Hereafter, we discuss the details of each period of the early universe.

%%%%%%%%%%%%%%%%%%%%%%%%%%%
%     Leptogenesis in the scotogenic model
%%%%%%%%%%%%%%%%%%%%%%%%%%%
\subsection{Leptogenesis in the scotogenic model}
\label{subsec:leptogenesis}

The baryon asymmetry of the universe is provided by the thermal leptogenesis scenario in our model.
In leptogeneis scenario, the lepton asymmetry is firstly generated by the decays of the right-handed neutrinos.
The generated lepton asymmetry depends on the following asymmetry parameters~\cite{Covi:1996wh}
\begin{align}
\label{eq:asymmetry-parameter}
   \epsilon_i =
   &\,\frac{\sum_\alpha \left[ \Gamma(N_i \to L_\alpha \eta) - \Gamma(N_i \to \bar{L}_\alpha \eta^\dag) \right]}{\sum_\alpha \left[ \Gamma(N_i \to L_\alpha \eta) + \Gamma(N_i \to \bar{L}_\alpha \eta^\dag) \right]} \nonumber \\
   =&\, \frac{1}{8\pi} \frac{1}{(h^\dag h)_{ii}} \sum_{j \neq i} {\rm Im}\left[ \left\{ \left( h^\dag h \right)_{ij} \right\}^2 \right] F(r_{ji},\eta_i)~, \\
   F(x,y) =
   &\,\sqrt{x} \left[ 1 + \frac{1+x-2y}{(1-y)^2} \ln \left( \frac{x-y^2}{1+x-2y} \right) - \frac{1}{x-1} (1-y)^2 \right]~.
\end{align}
Here we denote that the Dirac Yukawa coupling $h$ shown in Eq.~\eqref{eq:lagrangial}, $\eta_i \equiv m_\eta^2/M_i^2$, and $r_{ji} \equiv M_j^2/M_i^2$, and the function $F(x,y)$ comes from 
both the one-loop vertex contribution and the self-energy 
contribution. 

For the case that the masses of the right-handed neutrinos are
hierarchical, the lepton asymmetry is produced dominantly by the decays of the lightest right-handed neutrino $N_1$, and thus the baryon to photon number ratio is approximately given by~\cite{Buchmuller:2004nz}
\begin{align}
\label{eq:eta_B}
   \eta_B \approx - 0.01 \epsilon_1 \kappa_1~,
\end{align}
where $\kappa_1$ is the efficiency factor that presents the wash-out 
of the generated lepton asymmetry. This efficiency factor is calculated by the decay parameter $K_1 \equiv \Gamma_1 / H(T=M_1)$~\cite{Hugle:2018qbw} defined as 
\begin{align}
\label{eq:decay-para}
   K_1 &= 
   \frac{2 \pi^2}{\lambda_8} \xi_1 \sqrt{\frac{45}{64 \pi^5 g_*}} \frac{M_{\rm Pl}}{v^2} ~\widetilde{m}_{11}~(1 - \eta_1)^2 \nonumber \\
   &\simeq 15 \cdot \frac{10^{-7}}{\lambda_8} \left( \frac{-10}{\ln{\left(\eta_1\right)}} \right) \frac{\widetilde{m}_{11}}{10^{-10}\,{\rm eV}}~,
\end{align}
where $g_*$ stands for the effective number of relativistic degree of freedom, $M_{\rm Pl}$ does the Planck mass, and $\displaystyle \widetilde{m} \equiv R D_\nu R^\dag$.
As shown in Eq.~\eqref{eq:decay-para}, the decay parameter is much larger than 1 for the parameter region where we mainly focus and investigate. Thus in our scenario, the lepton asymmetry is generated via the strong wash-out regime.
For the large value of $K_1$, the efficiency factor can be approximated by~\cite{Buchmuller:2004nz}
\begin{align}
\label{eq:efficiency}
    \kappa_1(K_1) =
    \frac{1}{1.2 K_1 [\ln K_1]^{0.8}}~.
\end{align}

The asymmetry $n_{\Delta_\eta} = n_\eta - n_{\eta^\dag}$ makes the reaction 
rate of $\eta \eta \to H H$ larger than that of $\eta^\dag \eta^\dag \to 
H^\dag H^\dag$. 
The coupling constant $\lambda_8$ should be small so as that the mediator 
asymmetry is relayed to the DM asymmetry. 
In the non-relativistic regime of the mediator, the number densities 
$n_\eta$ and $n_{\eta^\dag}$ exponentially decay with temperature cooling. 
This implies that, if $(\sigma v_{\rm rel})_{\eta \eta \to H H}^{}\, n_{\eta^0} < H (T)$ 
in the relativistic regime, it holds in all of the regimes. 
Here $H(T)$ denotes the Hubble parameter. 
The annihilation cross section is roughly estimated in the relativistic regime by 
\begin{align}
    (\sigma v_{\rm rel})_{\eta \eta \to H H} =
    (\sigma v_{\rm rel})_{\eta^\dag \eta^\dag \to H^\dag H^\dag} =
    \frac{3 \lambda_8^2}{128 \pi T^2}~.
\label{Eq:relativisticXS}
\end{align}
The requirement $(\sigma v_{\rm rel})_{\eta \eta \to H H}^{}\, n_{\eta^0} < H (T)$ 
finds the following constraint; 
\begin{align}
\label{eq:lambda8-const}
    \lambda_8 < 3.9 \times 10^{-8} \sqrt{ \frac{T}{{\rm GeV}} }~.
\end{align}
The relativistic cross section in Eq.~\eqref{Eq:relativisticXS} is available at 
$T \gtrsim m_\eta$, and takes maximal for $T=m_\eta$. Most conservative 
constraint is found for $T=m_\eta$, e.g., $\lambda_8 < 3.9 \times 10^{-6}$ 
for $m_\eta = 10\,\text{TeV}$.

%%%%%%%%%%%%%%%%%%%%%%%%%%%
%     Mediator Annihilation and Dark Matter Production
%%%%%%%%%%%%%%%%%%%%%%%%%%%
\subsection{Mediator Annihilation and Dark Matter Production}
\label{subsec:mediator-DM}

The asymmetry of the mediator is produced through the CP-violating decays of the right-handed neutrinos. However, the mediator stays in thermal equilibrium in the early universe, and the asymmetry of $\eta$ and $\eta^\dag$ is much smaller than their number density, thus $n_{\Delta \eta}^{} \ll n_\eta, n_{\eta^\dag}$.
As the temperature of the universe falls below the mediator mass, the mediators annihilate into the weak and hypercharge gauge bosons.
After the annihilation of the mediators and satisfying $n_\eta \ll n_{\eta^\dag} \approx n_{\Delta \eta}$, the asymmetric component of the mediators decays into the DM, and the mediator asymmetry converts into the DM density.

For verifying whether the symmetric component of the mediator sufficiently annihilates, we evaluate the relic abundance of the mediator, assuming that the mediator has no asymmetry.
The relic density of the mediator after the annihilation process gets out of equilibrium can be calculated by~\cite{Kolb:1990vq}
\begin{align}
\label{eq:abundance}
   Y_{\eta,\infty} 
     \equiv \frac{n_{\eta,\infty}}{s}
     = 2 \times \frac{3.80\,x_{\mathrm{f}}}{\left(g_{* s} / g_{*}^{1 / 2}\right) M_{\rm Pl} m_{\eta} \braket{\sigma_{\rm g} v_{\rm rel}}}~,
\end{align}
where $s$ stands for the entropy density, $g_{*s}$ does the total relativistic degrees of freedom for entropy, and $\braket{\sigma_{\rm g} v_{\rm rel}}$ does the thermally averaged cross section of the mediator annihilation through the gauge interaction.
The factor of $2$ comes from the sum of the densities of $\eta^0$ and $\eta^+$.
In Eq.~\eqref{eq:abundance}, the ratio of the freeze-out temperature to the mediator mass, $x_{\rm f}$, is evaluated by
\begin{align}
\label{eq:xf}
    x_{\rm f} \equiv \frac{m_\eta}{T_{\rm f}} =
    &\ln \left[ 0.038 \left( g / g_{*}^{1/2} \right) M_{\rm Pl} m_{\eta} \braket{\sigma_{\rm g} v_{\rm rel}} \right] \notag \\
    &- \frac{1}{2} 
    \ln \left\{ \ln \left[0.038 \left(g / g_{*}^{1 / 2}\right) M_{\rm Pl} m_{\eta} \braket{\sigma_{\rm g} v_{\rm rel}} \right]\right\}~,
\end{align}
where $g$ is the internal degrees of freedom. 
Here we approximate the thermally averaged annihilation cross section by its non-relativistic limit as follows:
\begin{align}
   \braket{\sigma_{\rm g} v_{\rm rel}} \simeq
    \frac{ (g_1)^4 + 6 \cdot (g_1 g_2)^2 + 3 \cdot (g_2)^4}{256 \pi {m_\eta}^2 }~,
\end{align}
where $g_1$ and $g_2$ stand for the gauge couplings of the hypercharge and weak gauge bosons, respectively.

For the sufficient annihilation of the symmetric component, $Y_{\eta,\infty}$ should be smaller than the ratio of the mediator asymmetric component to the entropy density
as $\displaystyle Y_{\eta,\infty} < Y_{\Delta \eta}$. Note that 
we emphasize that our estimation above is conservative.
Assuming the existence of the asymmetric component, the number density of the annihilation partner is larger than that in no asymmetry case from the viewpoint of $\eta$. Therefore the annihilation of the mediator and decrease of the density of $\eta$ proceed more effectively, and the relation, $n_{\eta} \ll n_{\eta^\dag} \approx n_{\Delta \eta}$, can be realized more easily.

After the pair annihilation of the mediator, the asymmetric component decays into the DM and SM Higgs boson as $\eta^\dag \to H^\dag \sigma$. 
The decay width of the mediator is given by
\begin{align}
\label{eq:decay-width}
   \Gamma_{\rm decay} =
     \frac{\mu^2}{16 \pi m_\eta}~,
\end{align}
and the temperature when the mediator decay gets active, $T_{\rm decay}$, is obtained by $\Gamma_{\rm decay} = H(T_{\rm decay})$.
For the successful coincidence, the mediator should not decay before the annihilation of the symmetric component, and $T_{\rm decay}$ should satisfy $T_{\rm f} > T_{\rm decay}$.
From this condition, there is an upper limit on the scalar three-point coupling $\mu$.
On the other hand, the mediator decay during or after the BBN is cosmologically dangerous, and thus we request that the mediator decay starts by the BBN era, and the scalar three-point coupling satisfies $\Gamma_{\rm decay} > H(T_{\rm BBN})$ with $T_{\rm BBN} \simeq 1\,{\rm MeV}$ being the temperature when the BBN begins.
These two conditions give the following upper and lower bounds on the scalar three-point coupling:
\begin{align}
     8.4 \times 10^{-12}\, \frac{T_{\rm BBN}}{1\,{\rm MeV}} \sqrt{\frac{m_\eta}{\rm GeV}}
     <
     \frac{\mu}{\rm GeV}
     <
     8.4 \times 10^{-9}\, \frac{T_{\rm decay}}{\rm GeV} \sqrt{\frac{m_\eta}{\rm GeV}}~.
\end{align}

%We estimate the coupling constant $\mu$ from the decouple condition $H(T_{BBN}) < \Gamma_{\eta \rightarrow \sigma H } < H (T_{dec})$. The decay rate of $\eta$ is
%\begin{align}
%     \Gamma_{ \eta \rightarrow \sigma H }
%     =
%     \frac{{\mu}^2}{16 \pi m_\eta},
%\end{align}
%we have
%\begin{align}
%     8.41 \times 10^{-12} \, \text{[Gev]} \sqrt{\frac{m_\eta}{\text{[Gev]}}}
%     <
%     \mu
%     <
%     8.41 \times 10^{-9} \times T_{dec} \, \text{[Gev]} \sqrt{\frac{m_\eta}{\text{[Gev]}}}.
%\end{align}

%%%%%%%%%%%%%%%%%%%%%%%%%%%
%     Cosmological Constraints     %
%%%%%%%%%%%%%%%%%%%%%%%%%%%
\subsection{Cosmological Constraints on Asymmetric Mediator}
\label{subsec:constraint}

In the era of the BBN, a part of the DMs, which are produced by late time decays of the asymmetric mediators, can be relativistic.\,\footnote{
For $(m_\eta, m_\sigma) = (10^4\,{\rm GeV}, 5\,{\rm GeV})$ and $T_{\rm decay} = 1\,{\rm GeV}$, the momentum of the dark matter is given by $|\bm{p}_\sigma(t_{\rm BBN})| = m_\eta a(t_{\rm decay}) / 2 a(t_{\rm BBN}) = 0.5\,{\rm GeV}$.
Therefore, there is a possibility that a part of the DMs which decay for $T < T_{\rm decay}$ is relativistic.
}
Such a relativistic DM contributes to the expansion of the universe in the BBN and alter the BBN prediction.
In this subsection, we estimate this contribution by the DM and confirm that the DM in our model avoids the constraint from the BBN.

If the DM is relativistic in the BBN era, the energy density of the DM is given by
\begin{align}
\label{eq:Edensity-BBN}
   \rho_\sigma(t_{\rm BBN}) &=
   E_\sigma(t_{\rm BBN}) n_\sigma(t_{\rm BBN}) \nonumber \\
   &\simeq \frac{m_\eta}{2} \left( \frac{a(t_{\rm decay})}{a(t_{\rm BBN})} \right) \cdot n_\sigma(t_0) \left( \frac{a(t_0)}{a(t_{\rm BBN})} \right)^3 \nonumber \\
   &= 10^{25}\,{\rm GeV/cm^3} \cdot \frac{m_\eta}{10\,{\rm TeV}} \frac{a(t_{\rm decay})/a(t_{\rm BBN})}{10^{-3}} \nonumber \\
   &\qquad \quad \times \left( \frac{1~{\rm GeV}}{m_\sigma} \frac{\rho_\sigma(t_0)}{10^3~{\rm eV/cm^3}} \right) \left( \frac{a(t_0)/a(t_{\rm BBN})}{10^{10}} \right)^3~,
\end{align}
where $t_{\rm BBN}$, $t_{\rm decay}$, and $t_0$ are the times at the BBN, decays of the mediator, and present, respectively.
In Eq.~\eqref{eq:Edensity-BBN}, $\rho_\sigma$ and $n_\sigma$ stand for the energy and number densities of $\sigma$, respectively, and $E_\sigma$ does the energy of $\sigma$.
The deviation of the effective number of neutrino species, $\Delta N_{\rm eff} \equiv N_{\rm eff} - N_{\rm eff}^{\rm SM}$, is given by
\begin{align}
\label{eq:DNeff-BBN}
    \Delta N_{\rm eff} |_{\rm BBN} &=
    \frac{8}{7} \left( \frac{11}{4} \right)^{\frac{4}{3}} \frac{\rho_\sigma(t_{\rm BBN})}{\rho_\gamma(t_{\rm BBN})} \nonumber \\
    &= 2.7 \times 10^{-4} \cdot \frac{\rho_\sigma(t_{\rm BBN})}{10^{25}\,{\rm GeV/cm^3}} \frac{10^{28}\,{\rm GeV/cm^3}}{\rho_\gamma(t_{\rm BBN})}~,
\end{align}
with $\rho_\gamma$ being the energy density of the photon.
The contribution to the effective number of neutrino species by the DM in this model is much smaller than the SM prediction: $N_{\rm eff}^{SM} \simeq 3.044$~\cite{deSalas:2016ztq,Akita:2020szl,Froustey:2020mcq,Bennett:2020zkv}, and therefore the constraint on the DM from $N_{\rm eff}$ in the BBN era can be negligible.

%%%%%%%%%%%%%%%%%%%%%%%%%%%%%%%%%%%%%%%%%%
%     Result
%%%%%%%%%%%%%%%%%%%%%%%%%%%%%%%%%%%%%%%%%%
\section{Results}
\label{sec:result}
So far we have discussed the leptogenesis, annihilation of the symmetric component of the mediator, and DM production in the scotogenic model. In this section, we discuss the parameter region where our model realizes the coincidence between the observed baryon asymmetry and DM relic density without conflicting with the results of neutrino oscillation measurements.
In the following results, the neutrino oscillation parameters, such as the three mixing angles and two squared mass differences, and Dirac CP phase are fixed to be their best-fit values~\cite{Esteban:2020cvm}.
The two Majorana CP phases are fixed to be zero as a reference.
The three complex angles in the complex orthogonal matrix $R$ are varied as $10^{-10} < \left|\omega_i\right| < 1$ and $- \pi < {\rm arg} \left(\omega_i \right)< \pi$ ($i=1,2,3$).
The masses of the lightest active neutrino and mediator field are fixed to be $m_1 = 10^{-10}\,{\rm eV}$ and $\eta_1 \equiv m_\eta^2/M_1^2 = 10^{-6}$, respectively.
The mass hierarchy of the right-handed neutrinos are assumed to be $M_2 / M_1 = M_3 / M_2 = 1.5$.
As a reference, the value of $\lambda_8$ is fixed to be $\lambda_8 = 10^{-7}$ and $10^{-5}$.

%
%%% Figure %%%%%%%%%%%%%%%%%%%%%%%%%%%%%%%%%%%
\begin{figure}[t]
\centering
\includegraphics[width=0.49\textwidth]{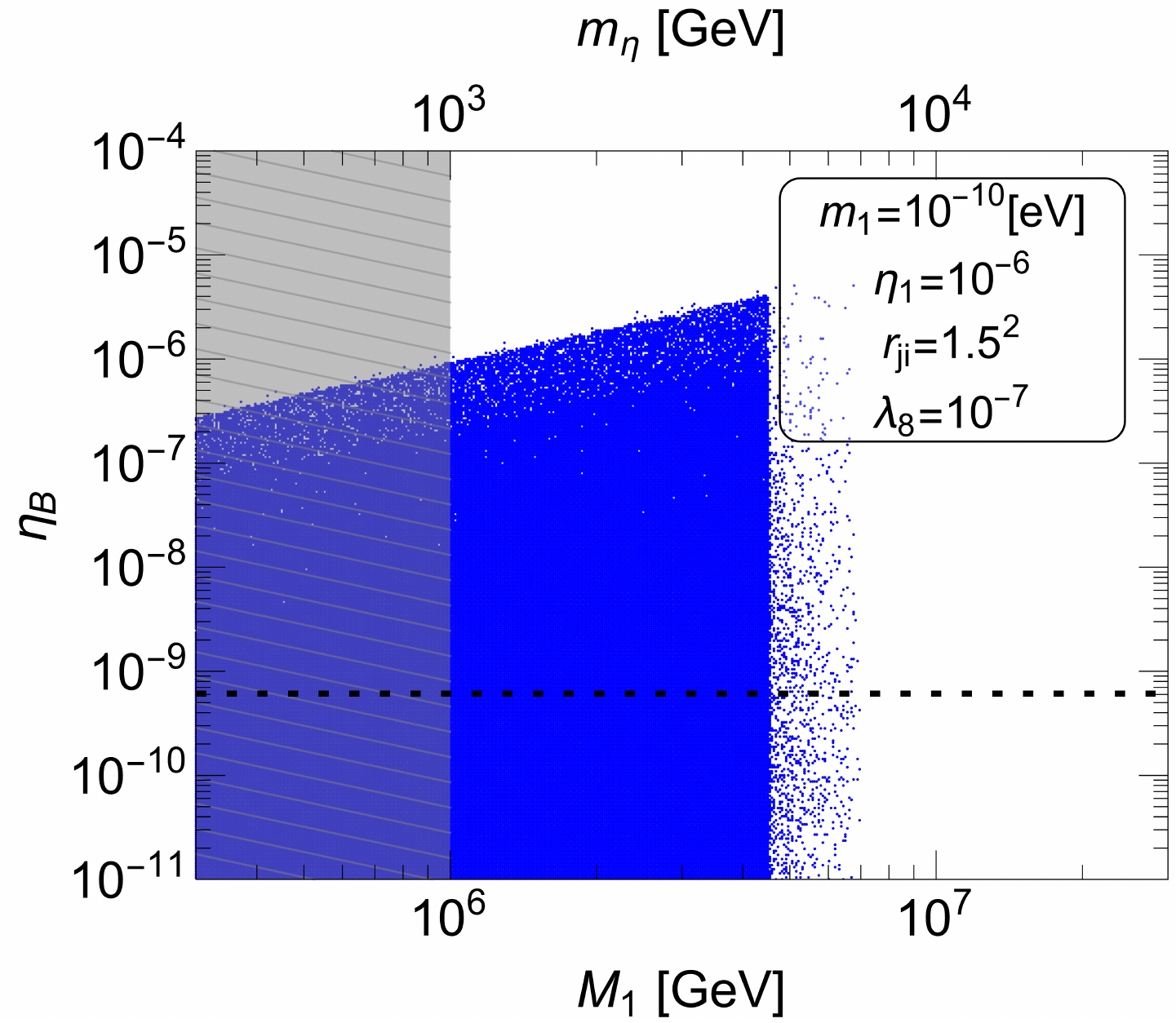}
\includegraphics[width=0.49\textwidth]{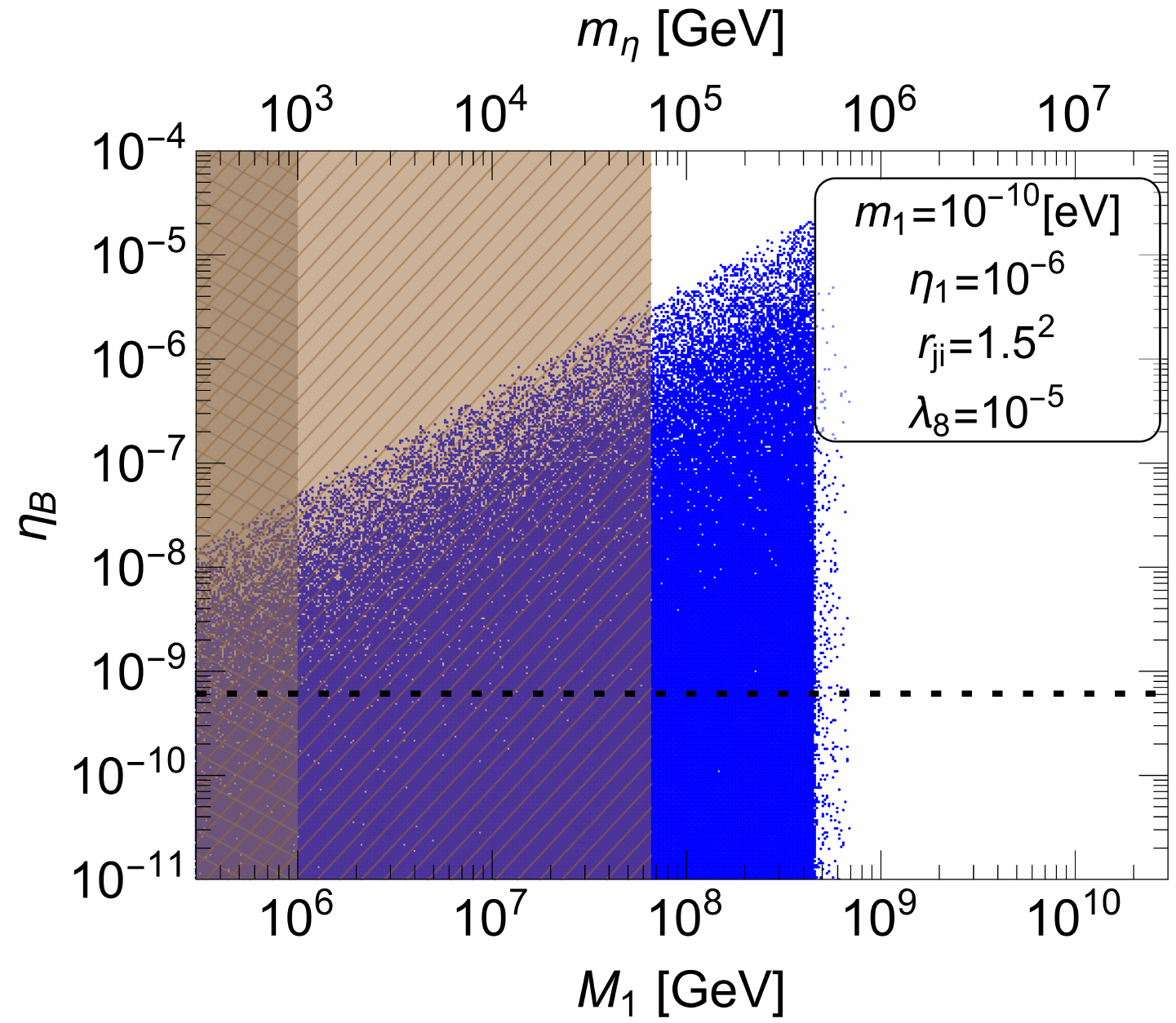}
\caption{
The baryon-to-photon ratio as a function of the mass of the lightest right-handed neutrino. 
All the scattered points avoid the constraint on the sum of the active neutrino masses and triviality bounds for the neutrino Yukawa couplings.
Moreover, we focus on the strong washout regime, and all the points satisfy $K_1 > 10$.
The black dotted line shows the observed baryon-to-photon ratio:~$\eta_B^{\rm obs} = 6.1 \times 10^{-10}$.
In the gray shaded region, the mediator is lighter than $1$\,TeV, and we do not consider such a region to avoid the severe collider bounds.  
The brown shaded region conflicts with the requirement in Eq.~\eqref{eq:lambda8-const}.
}
\label{fig:etaB-M1}
\end{figure}
%%%%%%%%%%%%%%%%%%%%%%%%%%%%%%%%%%%%%%%%%%%%
%
In Fig.~\ref{fig:etaB-M1}, we show the scatter plots of the baryon-to-photon ratio versus the mass of the lightest right-handed neutrino.
All the scattered points in Figs.~\ref{fig:etaB-M1} avoid the constraint on the sum of the active neutrino masses and triviality bounds for the neutrino Yukawa couplings, that is, $\sum_{i=1}^3 m_i < 0.16$\,eV (95\% C.L.)~\cite{Ivanov:2019hqk} and $\displaystyle\left|h_{\alpha i}\right|< 1$ for all $\alpha$ and $i$.
Moreover, we focus on the strong wash-out regime, and all the points satisfy $K_1 > 10$.
Therefore, the approximate formula of the efficiency factor in Eq.~\eqref{eq:efficiency} is valid.
The black dotted line shows the observed baryon-to-photon ratio:~$\eta_B^{\rm obs} = 6.1 \times 10^{-10}$.
In the gray shaded region, the mediator is lighter than $1$\,TeV, and we do not consider such a region to avoid the severe collider bounds.
The brown shaded region conflicts with the requirement in Eq.~\eqref{eq:lambda8-const}.

As shown in Fig.~\ref{fig:etaB-M1}, the observed baryon asymmetry can be generated in broad range of the lightest right-handed neutrino mass.
The large-$\eta_B$ boundary of the scattered region is determined by the active neutrino masses.
For a fixed $M_1$ (and $m_\eta$), too large Yukawa couplings cannot realize the light masses of the active neutrinos.
There is, therefore, a maximal value of the baryon-to-photon ratio. The heavy-$M_1$ boundary is determined by the triviality bound of the neutrino Yukawa couplings.
From Eq.~\eqref{eq:Mnu-approx}, the masses of the active neutrinos are obtained as
\begin{align}
    \mathcal{M}_\nu \simeq
    0.05\,{\rm eV} \cdot \frac{\lambda_8}{10^{-7}}\cdot \frac{h_{\alpha i} h_{\beta i}}{1} \cdot \frac{5 \times 10^{6}\,{\rm GeV}}{M_1}~,
\end{align}
for $m_\eta / M_1 = 10^{-3}$.
Thus, for $\left|h_{\alpha i}\right| < 1$ and $\lambda_8 = 10^{-7}$, the lightest right-handed neutrino with a mass heavier than $5 \times 10^6$~GeV conflicts with the observed neutrino oscillation and measured mass squared difference.

%
%%% Figure %%%%%%%%%%%%%%%%%%%%%%%%%%%%%%%%%%%
\begin{figure}[t]
\centering
\includegraphics[width=0.8\textwidth]{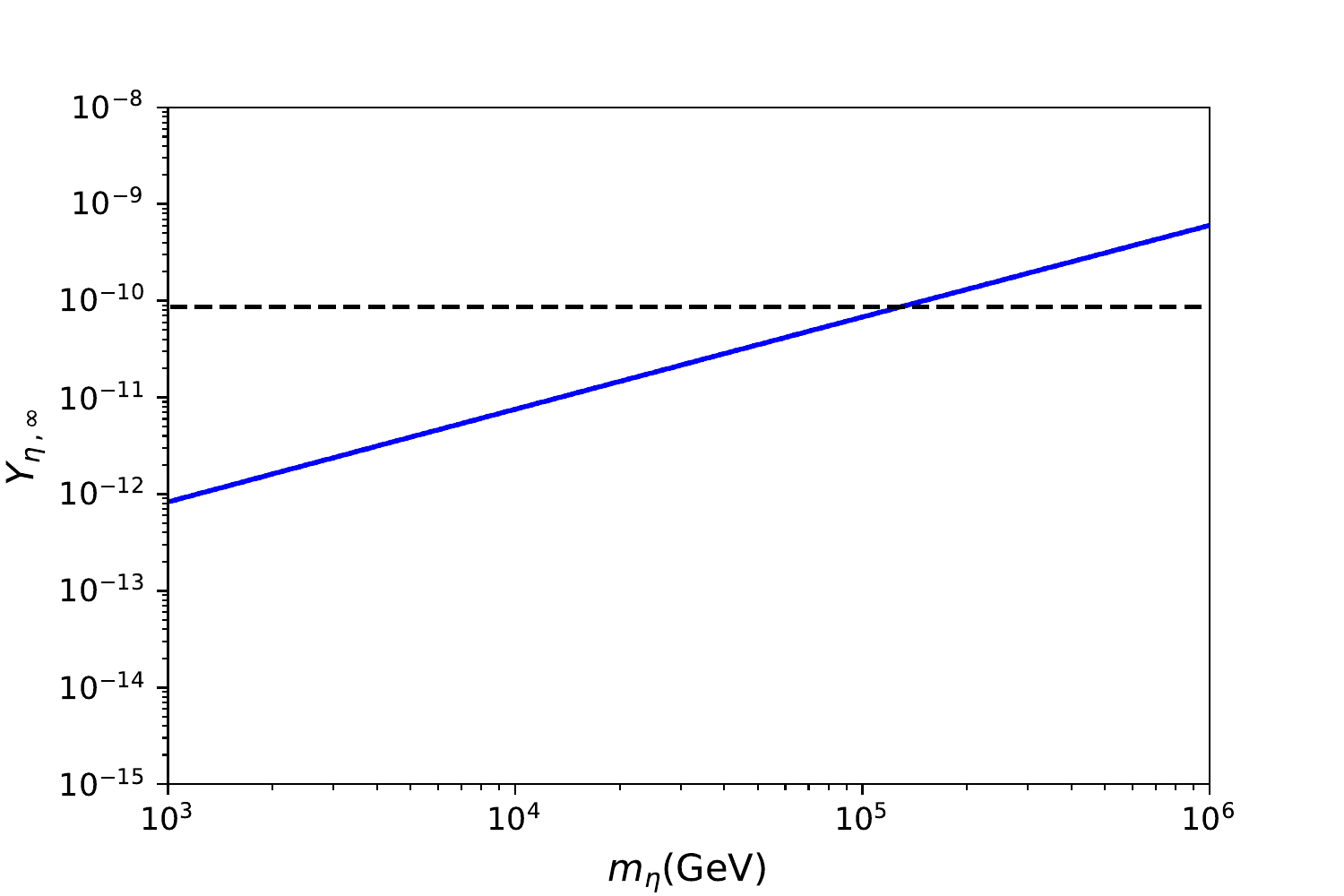}
\caption{
The relic abundance of $\eta$ after freeze-out, $Y_{\eta,\infty}$, as a function of the mediator mass.
The black dashed line corresponds to the observed baryon asymmetry, $Y_B = 8.66 \times 10^{-11}$.
}
\label{fig:abundance}
\end{figure}
%%%%%%%%%%%%%%%%%%%%%%%%%%%%%%%%%%%%%%%%%%%%
%

In Fig.~\ref{fig:abundance}, we show the relic abundance of $\eta$ after freeze-out as a function of the mediator mass. 
The black dashed line corresponds to the observed baryon asymmetry, $Y_B = 8.66 \times 10^{-11}$. 
From Fig.~\ref{fig:abundance}, we find that the mediator sufficiently annihilates, and hence the relation, $Y_{\eta,\infty} < Y_B$, is satisfied for $m_\eta \lesssim 10^{5}\,{\rm GeV}$.
In addition, if the mediator is heavier than the electroweak scale as $m_\eta \gg 10^2$\,GeV, our model can escape from severe collider bounds.

From Eq.~\eqref{eq:xf}, the freeze-out temperature is roughly obtained as $T_{\rm f} \sim m_\eta/22$.
The viable parameter region of the scalar three-point coupling is obtained as follows:
\begin{align}
   8.4 \times 10^{-12}\, \sqrt{\frac{m_\eta}{\rm Gev}}
   <
   \frac{\mu}{\rm GeV}
   <
   3.8 \times 10^{-10}\, \frac{T_{\rm decay}}{m_\eta/22} {\left(\frac{m_\eta}{\rm GeV}\right)}^{\frac{3}{2}}~.
\end{align}
According to the above results, our model can realizes the coincidence between the number densities of baryon and DM is realized as $Y_B \sim Y_L \simeq Y_{\Delta \eta} \simeq Y_{\rm DM}$ for the TeV scale mediator ($1\,{\rm TeV} \lesssim m_\eta \lesssim 10^2\,{\rm TeV}$) and the small scalar four and three-point couplings ($\lambda_8 \lesssim 10^{-8} \sqrt{m_\eta/{\rm GeV}},~10^{-11}\,{\rm GeV}\sqrt{m_\eta/{\rm GeV}} \lesssim \mu \lesssim 10^{-10}\,{\rm GeV} (m_\eta/{\rm GeV})^{3/2}$).

%%%%%%%%%%%%%%%%%%%%%%%%%%%%%%%%%%%%%%%%%%
%     Summary and Discussion
%%%%%%%%%%%%%%%%%%%%%%%%%%%%%%%%%%%%%%%%%%
\section{Summary and discussion}
\label{sec:summary}

The Scotogenic model is an excellent extension of the standard model to explain not only the light neutrino masses, mixing, and also DM. 
The purpose of this article is to combine the idea of ADM with the Scotogenic model to construct a model that simultaneously explains the relationship between the long-standing problems of the SM, namely, the neutrino mass and mixing, DM, and BAU.
Particularly, we focus on the coincidence between the observed energy densities of the DM and baryon as $\Omega_{\rm DM} / \Omega_B \approx 5$.
In this paper, we consider an extended model of the scotogenic model with a $Z_2$ odd singlet scalar field $\sigma$ that plays a role of DM. 
The $Z_2$ odd SU(2)$_L$ doublet scalar field $\eta$ is the mediator field, and it decays into $\sigma$. 
Since the mediator $\eta$ is simultaneously produced with the SM lepton $L_\alpha$ by the decay of the right-handed neutrino $N_i$, the asymmetry of the mediator, $n_{\Delta \eta}$, is the exactly same as that of the lepton, $n_{\Delta L}$. 
After the annihilation of the symmetric component of the mediators preserving the asymmetric component $n_{\Delta \eta}$, $n_\eta$ becomes much smaller than $n_{\eta^\dag}$ and $n_\eta \ll n_{\eta^\dag} \approx n_{\Delta \eta}$. 
Subsequently, the mediator decays into the DM, and the mediator asymmetry converts into the DM number density.
In this way, the DM number density can be related to the lepton asymmetry and baryon asymmetry of the universe generated by the leptogenesis scenario.

For the successful coincidence between the number densities of the lepton 
and mediator asymmetries, the CP violating annihilation, 
the interaction rates of 
$\eta \eta \to H H$ and $\eta^\dag \eta^\dag \to H^\dag H^\dag$ are required to be inactive, 
$\Gamma_{ \eta \eta \rightarrow H H (\eta^\dag \eta^\dag \to H^\dag H^\dag) } < H(T)$, because the unbalance rate between $\eta \eta \to H H$ and $\eta^\dag \eta^\dag \to H^\dag H^\dag$ 
distort the mediator asymmetry.
It requires that the scalar four-point coupling $\lambda_8$ should be small, though too small $\lambda_8$ spoils the radiative generation of the neutrino masses.
From this condition requirement, the scalar four-point coupling should satisfy $\lambda_8 < 3.9 \times 10^{-8} \times \sqrt{{m_\eta}/{\rm GeV}}$.
We calculate the baryon to photon number ratio $\eta_B$ taking $ \lambda_8 = 10^{-7}$ and $10^{-5}$ as reference values. 
Then we find that there exist successful parameters in range $10^{3}~{\rm GeV} < m_\eta < 10^{6}~{\rm GeV}$ that satisfies the observed value $\eta_B \simeq 6.1 \times 10^{-10}$.
Moreover, in order to match the number densities of baryons and DM, we require  $|Y_{\eta+\eta^\dag}| \simeq |Y_{\Delta \eta}|$, as DM is generated from $\eta$ and $\eta^\dagger$
decay. It means the symmetric component between $\eta$
and $\eta^\dagger$ annihilate completely.
From this condition, we found that the mediator mass should be lighter than $10^5$\,GeV.

In conclusion, the successful coincidence can be realized for $10^{3}\,{\rm GeV} \lesssim m_\eta \lesssim 10^{5}\,{\rm GeV}$, $\lambda_8 \lesssim 10^{-8} \sqrt{m_\eta/{\rm GeV}}$, and $~10^{-11}\,{\rm GeV}\sqrt{m_\eta/{\rm GeV}} \lesssim \mu \lesssim 10^{-10}\,{\rm GeV} (m_\eta/{\rm GeV})^{3/2}$.
There is a possibility to observe the mediator in this model in accelerator experiments in the near future.
This study will be shown elsewhere.

%%%%%%%%%%%%%%%%%%%%%%%%%%%%%%%%%%%%%%%%%%
%     Acknowledgments
%%%%%%%%%%%%%%%%%%%%%%%%%%%%%%%%%%%%%%%%%%
\section*{Acknowledgments}
The authors would like to thank Hiroaki Sugiyama for useful comments.
This work is supported in part by the Grant-in-Aid for Research Activity Start-up (No.21K20365 [KA]) and Scientific Research (No.18H01210 [KA, JS, and YT], 
22K03638, 22K03602, JP20H05852 [MY]) and MEXT KAKENHI Grant (No.18H05543 [KA, JS, and YT]). This work was partly supported by MEXT Joint Usage/Research Center on Mathematics and Theoretical Physics JPMXP0619217849 [MY].

{\small
\bibliography{ref}

\providecommand{\href}[2]{#2}\begingroup\begin{thebibliography}{10}

\bibitem{Fukugita:1986hr}
M.~Fukugita and T.~Yanagida, ``{Baryogenesis Without Grand Unification},''
  \href{https://dx.doi.org/10.1016/0370-2693(86)91126-3}{Phys.\  Lett.\  B
  {\bfseries 174} (1986) 45--47}.

\bibitem{Planck:2018vyg}
{\bfseries Planck} Collaboration, ``{Planck 2018 results. VI. Cosmological
  parameters},'' \href{https://dx.doi.org/10.1051/0004-6361/201833910}{Astron.\
   Astrophys.\  {\bfseries 641} (2020) A6} {\ttfamily
  [\href{https://arxiv.org/abs/1807.06209}{arXiv:1807.06209}]}. [Erratum:
  Astron.Astrophys. 652, C4 (2021)].

\bibitem{Nussinov:1985xr}
S.~Nussinov, ``{TECHNOCOSMOLOGY: COULD A TECHNIBARYON EXCESS PROVIDE A
  'NATURAL' MISSING MASS CANDIDATE?}''
  \href{https://dx.doi.org/10.1016/0370-2693(85)90689-6}{Phys.\  Lett.\  B
  {\bfseries 165} (1985) 55--58}.

\bibitem{Barr:1990ca}
S.~M.~Barr, R.~S.~Chivukula, and E.~Farhi, ``{Electroweak Fermion Number
  Violation and the Production of Stable Particles in the Early Universe},''
  \href{https://dx.doi.org/10.1016/0370-2693(90)91661-T}{Phys.\  Lett.\  B
  {\bfseries 241} (1990) 387--391}.

\bibitem{Barr:1991qn}
S.~M.~Barr, ``{Baryogenesis, sphalerons and the cogeneration of dark matter},''
  \href{https://dx.doi.org/10.1103/PhysRevD.44.3062}{Phys.\  Rev.\  D
  {\bfseries 44} (1991) 3062--3066}.

\bibitem{Dodelson:1991iv}
S.~Dodelson, B.~R.~Greene, and L.~M.~Widrow, ``{Baryogenesis, dark matter and
  the width of the Z},''
  \href{https://dx.doi.org/10.1016/0550-3213(92)90328-9}{Nucl.\  Phys.\  B
  {\bfseries 372} (1992) 467--493}.

\bibitem{Kaplan:1991ah}
D.~B.~Kaplan, ``{A Single explanation for both the baryon and dark matter
  densities},'' \href{https://dx.doi.org/10.1103/PhysRevLett.68.741}{Phys.\
  Rev.\  Lett.\  {\bfseries 68} (1992) 741--743}.

\bibitem{Kuzmin:1996he}
S.~J.~Ball and Y.~A.~Kamyshkov, eds., ``{A Simultaneous solution to
  baryogenesis and dark matter problems},''
  \href{https://dx.doi.org/10.1134/1.953070}{Phys.\  Part.\  Nucl.\  {\bfseries
  29} (1998) 257--265} {\ttfamily
  [\href{https://arxiv.org/abs/hep-ph/9701269}{hep-ph/9701269}]}.

\bibitem{Foot:2003jt}
R.~Foot and R.~R.~Volkas, ``{Was ordinary matter synthesized from mirror
  matter? An Attempt to explain why Omega(Baryon) approximately equal to 0.2
  Omega(Dark)},'' \href{https://dx.doi.org/10.1103/PhysRevD.68.021304}{Phys.\
  Rev.\  D {\bfseries 68} (2003) 021304} {\ttfamily
  [\href{https://arxiv.org/abs/hep-ph/0304261}{hep-ph/0304261}]}.

\bibitem{Foot:2004pq}
R.~Foot and R.~R.~Volkas, ``{Explaining Omega(Baryon) approximately 0.2
  Omega(Dark) through the synthesis of ordinary matter from mirror matter: A
  More general analysis},''
  \href{https://dx.doi.org/10.1103/PhysRevD.69.123510}{Phys.\  Rev.\  D
  {\bfseries 69} (2004) 123510} {\ttfamily
  [\href{https://arxiv.org/abs/hep-ph/0402267}{hep-ph/0402267}]}.

\bibitem{Hooper:2004dc}
D.~Hooper, J.~March-Russell, and S.~M.~West, ``{Asymmetric sneutrino dark
  matter and the Omega(b) / Omega(DM) puzzle},''
  \href{https://dx.doi.org/10.1016/j.physletb.2004.11.047}{Phys.\  Lett.\  B
  {\bfseries 605} (2005) 228--236} {\ttfamily
  [\href{https://arxiv.org/abs/hep-ph/0410114}{hep-ph/0410114}]}.

\bibitem{Kitano:2004sv}
R.~Kitano and I.~Low, ``{Dark matter from baryon asymmetry},''
  \href{https://dx.doi.org/10.1103/PhysRevD.71.023510}{Phys.\  Rev.\  D
  {\bfseries 71} (2005) 023510} {\ttfamily
  [\href{https://arxiv.org/abs/hep-ph/0411133}{hep-ph/0411133}]}.

\bibitem{Gudnason:2006ug}
S.~B.~Gudnason, C.~Kouvaris, and F.~Sannino, ``{Towards working technicolor:
  Effective theories and dark matter},''
  \href{https://dx.doi.org/10.1103/PhysRevD.73.115003}{Phys.\  Rev.\  D
  {\bfseries 73} (2006) 115003} {\ttfamily
  [\href{https://arxiv.org/abs/hep-ph/0603014}{hep-ph/0603014}]}.

\bibitem{Kaplan:2009ag}
D.~E.~Kaplan, M.~A.~Luty, and K.~M.~Zurek, ``{Asymmetric Dark Matter},''
  \href{https://dx.doi.org/10.1103/PhysRevD.79.115016}{Phys.\  Rev.\  D
  {\bfseries 79} (2009) 115016} {\ttfamily
  [\href{https://arxiv.org/abs/0901.4117}{arXiv:0901.4117}]}.

\bibitem{Davoudiasl:2012uw}
H.~Davoudiasl and R.~N.~Mohapatra, ``{On Relating the Genesis of Cosmic Baryons
  and Dark Matter},''
  \href{https://dx.doi.org/10.1088/1367-2630/14/9/095011}{New J.\  Phys.\
  {\bfseries 14} (2012) 095011} {\ttfamily
  [\href{https://arxiv.org/abs/1203.1247}{arXiv:1203.1247}]}.

\bibitem{Petraki:2013wwa}
K.~Petraki and R.~R.~Volkas, ``{Review of asymmetric dark matter},''
  \href{https://dx.doi.org/10.1142/S0217751X13300287}{Int.\  J.\  Mod.\  Phys.\
   A {\bfseries 28} (2013) 1330028} {\ttfamily
  [\href{https://arxiv.org/abs/1305.4939}{arXiv:1305.4939}]}.

\bibitem{Zurek:2013wia}
K.~M.~Zurek, ``{Asymmetric Dark Matter: Theories, Signatures, and
  Constraints},''
  \href{https://dx.doi.org/10.1016/j.physrep.2013.12.001}{Phys.\  Rept.\
  {\bfseries 537} (2014) 91--121} {\ttfamily
  [\href{https://arxiv.org/abs/1308.0338}{arXiv:1308.0338}]}.

\bibitem{Ma:2006km}
E.~Ma, ``{Verifiable radiative seesaw mechanism of neutrino mass and dark
  matter},'' \href{https://dx.doi.org/10.1103/PhysRevD.73.077301}{Phys.\  Rev.\
   D {\bfseries 73} (2006) 077301} {\ttfamily
  [\href{https://arxiv.org/abs/hep-ph/0601225}{hep-ph/0601225}]}.

\bibitem{Minkowski:1977sc}
P.~Minkowski, ``{$\mu \to e\gamma$ at a Rate of One Out of $10^{9}$ Muon
  Decays?}'' \href{https://dx.doi.org/10.1016/0370-2693(77)90435-X}{Phys.\
  Lett.\  B {\bfseries 67} (1977) 421--428}.

\bibitem{Yanagida:1979as}
O.~Sawada and A.~Sugamoto, eds., ``{Horizontal gauge symmetry and masses of
  neutrinos},'' Conf.\  Proc.\  C {\bfseries 7902131} (1979) 95--99.

\bibitem{Gell-Mann:1979vob}
M.~Gell-Mann, P.~Ramond, and R.~Slansky, ``{Complex Spinors and Unified
  Theories},'' Conf.\  Proc.\  C {\bfseries 790927} (1979) 315--321 {\ttfamily
  [\href{https://arxiv.org/abs/1306.4669}{arXiv:1306.4669}]}.

\bibitem{Mohapatra:1979ia}
R.~N.~Mohapatra and G.~Senjanovic, ``{Neutrino Mass and Spontaneous Parity
  Nonconservation},''
  \href{https://dx.doi.org/10.1103/PhysRevLett.44.912}{Phys.\  Rev.\  Lett.\
  {\bfseries 44} (1980) 912}.

\bibitem{Goldman:1989nd}
I.~Goldman and S.~Nussinov, ``{Weakly Interacting Massive Particles and Neutron
  Stars},'' \href{https://dx.doi.org/10.1103/PhysRevD.40.3221}{Phys.\  Rev.\  D
  {\bfseries 40} (1989) 3221--3230}.

\bibitem{Gould:1989gw}
A.~Gould, B.~T.~Draine, R.~W.~Romani, and S.~Nussinov, ``{Neutron Stars:
  Graveyard of Charged Dark Matter},''
  \href{https://dx.doi.org/10.1016/0370-2693(90)91745-W}{Phys.\  Lett.\  B
  {\bfseries 238} (1990) 337--343}.

\bibitem{Kouvaris:2007ay}
C.~Kouvaris, ``{WIMP Annihilation and Cooling of Neutron Stars},''
  \href{https://dx.doi.org/10.1103/PhysRevD.77.023006}{Phys.\  Rev.\  D
  {\bfseries 77} (2008) 023006} {\ttfamily
  [\href{https://arxiv.org/abs/0708.2362}{arXiv:0708.2362}]}.

\bibitem{Kouvaris:2010vv}
C.~Kouvaris and P.~Tinyakov, ``{Can Neutron stars constrain Dark Matter?}''
  \href{https://dx.doi.org/10.1103/PhysRevD.82.063531}{Phys.\  Rev.\  D
  {\bfseries 82} (2010) 063531} {\ttfamily
  [\href{https://arxiv.org/abs/1004.0586}{arXiv:1004.0586}]}.

\bibitem{deLavallaz:2010wp}
A.~de~Lavallaz and M.~Fairbairn, ``{Neutron Stars as Dark Matter Probes},''
  \href{https://dx.doi.org/10.1103/PhysRevD.81.123521}{Phys.\  Rev.\  D
  {\bfseries 81} (2010) 123521} {\ttfamily
  [\href{https://arxiv.org/abs/1004.0629}{arXiv:1004.0629}]}.

\bibitem{McDermott:2011jp}
S.~D.~McDermott, H.-B.~Yu, and K.~M.~Zurek, ``{Constraints on Scalar Asymmetric
  Dark Matter from Black Hole Formation in Neutron Stars},''
  \href{https://dx.doi.org/10.1103/PhysRevD.85.023519}{Phys.\  Rev.\  D
  {\bfseries 85} (2012) 023519} {\ttfamily
  [\href{https://arxiv.org/abs/1103.5472}{arXiv:1103.5472}]}.

\bibitem{Kouvaris:2011fi}
C.~Kouvaris and P.~Tinyakov, ``{Excluding Light Asymmetric Bosonic Dark
  Matter},'' \href{https://dx.doi.org/10.1103/PhysRevLett.107.091301}{Phys.\
  Rev.\  Lett.\  {\bfseries 107} (2011) 091301} {\ttfamily
  [\href{https://arxiv.org/abs/1104.0382}{arXiv:1104.0382}]}.

\bibitem{Guver:2012ba}
T.~G\"uver, A.~E.~Erkoca, M.~Hall~Reno, and I.~Sarcevic, ``{On the capture of
  dark matter by neutron stars},''
  \href{https://dx.doi.org/10.1088/1475-7516/2014/05/013}{JCAP {\bfseries 05}
  (2014) 013} {\ttfamily
  [\href{https://arxiv.org/abs/1201.2400}{arXiv:1201.2400}]}.

\bibitem{Bell:2013xk}
N.~F.~Bell, A.~Melatos, and K.~Petraki, ``{Realistic neutron star constraints
  on bosonic asymmetric dark matter},''
  \href{https://dx.doi.org/10.1103/PhysRevD.87.123507}{Phys.\  Rev.\  D
  {\bfseries 87} (2013) 123507} {\ttfamily
  [\href{https://arxiv.org/abs/1301.6811}{arXiv:1301.6811}]}.

\bibitem{Bramante:2013hn}
J.~Bramante, K.~Fukushima, and J.~Kumar, ``{Constraints on bosonic dark matter
  from observation of old neutron stars},''
  \href{https://dx.doi.org/10.1103/PhysRevD.87.055012}{Phys.\  Rev.\  D
  {\bfseries 87} (2013) 055012} {\ttfamily
  [\href{https://arxiv.org/abs/1301.0036}{arXiv:1301.0036}]}.

\bibitem{Bramante:2013nma}
J.~Bramante, K.~Fukushima, J.~Kumar, and E.~Stopnitzky, ``{Bounds on
  self-interacting fermion dark matter from observations of old neutron
  stars},'' \href{https://dx.doi.org/10.1103/PhysRevD.89.015010}{Phys.\  Rev.\
  D {\bfseries 89} (2014) 015010} {\ttfamily
  [\href{https://arxiv.org/abs/1310.3509}{arXiv:1310.3509}]}.

\bibitem{Kouvaris:2013kra}
C.~Kouvaris and P.~Tinyakov, ``{Growth of Black Holes in the interior of
  Rotating Neutron Stars},''
  \href{https://dx.doi.org/10.1103/PhysRevD.90.043512}{Phys.\  Rev.\  D
  {\bfseries 90} (2014) 043512} {\ttfamily
  [\href{https://arxiv.org/abs/1312.3764}{arXiv:1312.3764}]}.

\bibitem{Bramante:2014zca}
J.~Bramante and T.~Linden, ``{Detecting Dark Matter with Imploding Pulsars in
  the Galactic Center},''
  \href{https://dx.doi.org/10.1103/PhysRevLett.113.191301}{Phys.\  Rev.\
  Lett.\  {\bfseries 113} (2014) 191301} {\ttfamily
  [\href{https://arxiv.org/abs/1405.1031}{arXiv:1405.1031}]}.

\bibitem{Ilie:2020vec}
C.~Ilie, J.~Pilawa, and S.~Zhang, ``{Comment on \textquotedblleft{}Multiscatter
  stellar capture of dark matter\textquotedblright{}},''
  \href{https://dx.doi.org/10.1103/PhysRevD.102.048301}{Phys.\  Rev.\  D
  {\bfseries 102} (2020) 048301} {\ttfamily
  [\href{https://arxiv.org/abs/2005.05946}{arXiv:2005.05946}]}.

\bibitem{Kouvaris:2018wnh}
C.~Kouvaris, P.~Tinyakov, and M.~H.~G.~Tytgat, ``{NonPrimordial Solar Mass
  Black Holes},''
  \href{https://dx.doi.org/10.1103/PhysRevLett.121.221102}{Phys.\  Rev.\
  Lett.\  {\bfseries 121} (2018) 221102} {\ttfamily
  [\href{https://arxiv.org/abs/1804.06740}{arXiv:1804.06740}]}.

\bibitem{Gresham:2018rqo}
M.~I.~Gresham and K.~M.~Zurek, ``{Asymmetric Dark Stars and Neutron Star
  Stability},'' \href{https://dx.doi.org/10.1103/PhysRevD.99.083008}{Phys.\
  Rev.\  D {\bfseries 99} (2019) 083008} {\ttfamily
  [\href{https://arxiv.org/abs/1809.08254}{arXiv:1809.08254}]}.

\bibitem{Garani:2018kkd}
R.~Garani, Y.~Genolini, and T.~Hambye, ``{New Analysis of Neutron Star
  Constraints on Asymmetric Dark Matter},''
  \href{https://dx.doi.org/10.1088/1475-7516/2019/05/035}{JCAP {\bfseries 05}
  (2019) 035} {\ttfamily
  [\href{https://arxiv.org/abs/1812.08773}{arXiv:1812.08773}]}.

\bibitem{Borah:2018uci}
D.~Borah, A.~Dasgupta, and S.~K.~Kang, ``{TeV Scale Leptogenesis via Dark
  Sector Scatterings},''
  \href{https://dx.doi.org/10.1140/epjc/s10052-020-8052-1}{Eur.\  Phys.\  J.\
  C {\bfseries 80} (2020) 498} {\ttfamily
  [\href{https://arxiv.org/abs/1806.04689}{arXiv:1806.04689}]}.

\bibitem{Borah:2019epq}
D.~Borah, A.~Dasgupta, and S.~K.~Kang, ``{Two-component dark matter with
  cogenesis of the baryon asymmetry of the Universe},''
  \href{https://dx.doi.org/10.1103/PhysRevD.100.103502}{Phys.\  Rev.\  D
  {\bfseries 100} (2019) 103502} {\ttfamily
  [\href{https://arxiv.org/abs/1903.10516}{arXiv:1903.10516}]}.

\bibitem{Casas:2001sr}
J.~A.~Casas and A.~Ibarra, ``{Oscillating neutrinos and $\mu \to e, \gamma$},''
  \href{https://dx.doi.org/10.1016/S0550-3213(01)00475-8}{Nucl.\  Phys.\  B
  {\bfseries 618} (2001) 171--204} {\ttfamily
  [\href{https://arxiv.org/abs/hep-ph/0103065}{hep-ph/0103065}]}.

\bibitem{Hugle:2018qbw}
T.~Hugle, M.~Platscher, and K.~Schmitz, ``{Low-Scale Leptogenesis in the
  Scotogenic Neutrino Mass Model},''
  \href{https://dx.doi.org/10.1103/PhysRevD.98.023020}{Phys.\  Rev.\  D
  {\bfseries 98} (2018) 023020} {\ttfamily
  [\href{https://arxiv.org/abs/1804.09660}{arXiv:1804.09660}]}.

\bibitem{Pontecorvo:1967fh}
B.~Pontecorvo, ``{Neutrino Experiments and the Problem of Conservation of
  Leptonic Charge},'' Sov.\  Phys.\  JETP {\bfseries 26} (1968) 984--988.
[Zh. Eksp. Teor. Fiz.53,1717(1967)].
%%CITATION = SPHJA,26,984;%%.

\bibitem{Pontecorvo:1957cp}
B.~Pontecorvo, ``{Mesonium and anti-mesonium},'' Sov.\  Phys.\  JETP {\bfseries
  6} (1957) 429.
[Zh. Eksp. Teor. Fiz.33,549(1957)].
%%CITATION = SPHJA,6,429;%%.

\bibitem{Pontecorvo:1957qd}
B.~Pontecorvo, ``{Inverse beta processes and nonconservation of lepton
  charge},'' Sov.\  Phys.\  JETP {\bfseries 7} (1958) 172--173.
[Zh. Eksp. Teor. Fiz.34,247(1957)].
%%CITATION = SPHJA,7,172;%%.

\bibitem{Maki:1962mu}
Z.~Maki, M.~Nakagawa, and S.~Sakata, ``{Remarks on the unified model of
  elementary particles},''
\href{https://dx.doi.org/10.1143/PTP.28.870}{Prog.\  Theor.\  Phys.\
  {\bfseries 28} (1962) 870--880}.
%%CITATION = PTPKA,28,870;%%.

\bibitem{Zyla:2020zbs}
{\bfseries Particle Data Group} Collaboration, ``{Review of Particle
  Physics},'' \href{https://dx.doi.org/10.1093/ptep/ptaa104}{PTEP {\bfseries
  2020} (2020) 083C01}.

\bibitem{Covi:1996wh}
L.~Covi, E.~Roulet, and F.~Vissani, ``{CP violating decays in leptogenesis
  scenarios},'' \href{https://dx.doi.org/10.1016/0370-2693(96)00817-9}{Phys.\
  Lett.\  B {\bfseries 384} (1996) 169--174} {\ttfamily
  [\href{https://arxiv.org/abs/hep-ph/9605319}{hep-ph/9605319}]}.

\bibitem{Buchmuller:2004nz}
W.~Buchmuller, P.~Di~Bari, and M.~Plumacher, ``{Leptogenesis for
  pedestrians},'' \href{https://dx.doi.org/10.1016/j.aop.2004.02.003}{Annals
  Phys.\  {\bfseries 315} (2005) 305--351} {\ttfamily
  [\href{https://arxiv.org/abs/hep-ph/0401240}{hep-ph/0401240}]}.

\bibitem{Kolb:1990vq}
E.~W.~Kolb and M.~S.~Turner,
  \href{https://dx.doi.org/10.1201/9780429492860}{{\em {The Early Universe}}},
  vol.~69.
\newblock 1990.

\bibitem{deSalas:2016ztq}
P.~F.~de~Salas and S.~Pastor, ``{Relic neutrino decoupling with flavour
  oscillations revisited},''
  \href{https://dx.doi.org/10.1088/1475-7516/2016/07/051}{JCAP {\bfseries 07}
  (2016) 051} {\ttfamily
  [\href{https://arxiv.org/abs/1606.06986}{arXiv:1606.06986}]}.

\bibitem{Akita:2020szl}
K.~Akita and M.~Yamaguchi, ``{A precision calculation of relic neutrino
  decoupling},'' \href{https://dx.doi.org/10.1088/1475-7516/2020/08/012}{JCAP
  {\bfseries 08} (2020) 012} {\ttfamily
  [\href{https://arxiv.org/abs/2005.07047}{arXiv:2005.07047}]}.

\bibitem{Froustey:2020mcq}
J.~Froustey, C.~Pitrou, and M.~C.~Volpe, ``{Neutrino decoupling including
  flavour oscillations and primordial nucleosynthesis},''
  \href{https://dx.doi.org/10.1088/1475-7516/2020/12/015}{JCAP {\bfseries 12}
  (2020) 015} {\ttfamily
  [\href{https://arxiv.org/abs/2008.01074}{arXiv:2008.01074}]}.

\bibitem{Bennett:2020zkv}
J.~J.~Bennett, G.~Buldgen, P.~F.~De~Salas, M.~Drewes, \emph{et al}., ``{Towards
  a precision calculation of $N_{\rm eff}$ in the Standard Model II: Neutrino
  decoupling in the presence of flavour oscillations and finite-temperature
  QED},'' \href{https://dx.doi.org/10.1088/1475-7516/2021/04/073}{JCAP
  {\bfseries 04} (2021) 073} {\ttfamily
  [\href{https://arxiv.org/abs/2012.02726}{arXiv:2012.02726}]}.

\bibitem{Esteban:2020cvm}
I.~Esteban, M.~C.~Gonzalez-Garcia, M.~Maltoni, T.~Schwetz, and A.~Zhou, ``{The
  fate of hints: updated global analysis of three-flavor neutrino
  oscillations},'' \href{https://dx.doi.org/10.1007/JHEP09(2020)178}{JHEP
  {\bfseries 09} (2020) 178} {\ttfamily
  [\href{https://arxiv.org/abs/2007.14792}{arXiv:2007.14792}]}.

\bibitem{Ivanov:2019hqk}
M.~M.~Ivanov, M.~Simonovi\'c, and M.~Zaldarriaga, ``{Cosmological Parameters
  and Neutrino Masses from the Final Planck and Full-Shape BOSS Data},''
  \href{https://dx.doi.org/10.1103/PhysRevD.101.083504}{Phys.\  Rev.\  D
  {\bfseries 101} (2020) 083504} {\ttfamily
  [\href{https://arxiv.org/abs/1912.08208}{arXiv:1912.08208}]}.

\end{thebibliography}\endgroup
}

\end{document}